\documentclass[sigplan, nonacm]{acmart}

\usepackage{graphicx}
\usepackage{microtype}
\usepackage{enumitem}
\usepackage{listings}
\usepackage{float}
\usepackage{tikzit}
\usepackage[switch]{lineno}
\usepackage{subcaption}
\usepackage{mathtools}
\usepackage{tikz}
\usepackage{accents}
\usepackage{scalerel}
\usepackage{rotating}
\usepackage{stackengine,xcolor}
\usepackage{array}%
\usepackage{adjustbox}
\usepackage{wasysym}
\usepackage[outline]{contour}
\usepackage{comment}
\usepackage{xspace}
\usepackage{soul}
\usepackage{physics}
\usepackage[para]{footmisc}

\usetikzlibrary{decorations.pathreplacing,calc}

\acmJournal{PACMPL}
\renewcommand\footnotetextcopyrightpermission[1]{}
\setcopyright{none}

\makeatletter
\let\@authorsaddresses\@empty
\makeatother

\tikzstyle{rectangle}=[fill=white, draw=black, shape=rectangle]
\tikzstyle{rect-grey}=[fill={rgb,255: red,213; green,213; blue,213}, draw=black, shape=rectangle]
\tikzstyle{rect-grey-fixedwidth}=[fill={rgb,255: red,213; green,213; blue,213}, draw=black, shape=rectangle, style={align=center, text width=20mm}]
\tikzstyle{rect-grey-fixed}=[fill={rgb,255: red,213; green,213; blue,213}, draw=black, shape=rectangle, style={align=center, text width=20mm, minimum height = 1.1cm}]
\tikzstyle{inference}=[fill=inferenceCol, draw=black, shape=rectangle, style={align=center, text width=16mm, minimum height = 1.1cm}]
\tikzstyle{rect-dummy-fixed}=[draw=none, shape=rectangle, style={text width=60mm, minimum height = 1.1cm}]
\tikzstyle{rect-white}=[fill=white, draw=none, shape=rectangle]
\tikzstyle{rect-white-fixedheight}=[fill=white, draw=black, shape=rectangle, style={minimum height = 0.6cm, text height=1mm}]
\tikzstyle{abstract-model}=[fill=modelCol, draw=black, shape=rectangle, style={dotted-line, align=center,  minimum height = 1.2cm, minimum width = 3cm}]
\tikzstyle{abstract-inference}=[fill=inferenceCol, draw=black, shape=rectangle, style={dotted-line, align=center, minimum height = 1.3cm, minimum width = 2.8cm}]
\tikzstyle{abstract-inference-wide}=[fill=inferenceCol, draw=black, shape=rectangle, style={dotted-line, align=center,  minimum height = 1.3cm, minimum width = 3.2cm}]

\tikzstyle{square-abstract-model}=[fill=modelCol, draw=black, shape=rectangle, style={dotted-line, align=center,  minimum height = 0.4cm, minimum width = 0.4cm}]
\tikzstyle{square-abstract-inference}=[fill=inferenceCol, draw=black, shape=rectangle, style={dotted-line, align=center,  minimum height = 0.4cm, minimum width = 0.4cm}]
\tikzstyle{square-inference}=[fill=inferenceCol, draw=black, shape=rectangle, style={align=center,  minimum height = 0.4cm, minimum width = 0.4cm}]
\tikzstyle{square-white}=[fill=white, draw=black, shape=rectangle, style={align=center,  minimum height = 0.4cm, minimum width = 0.4cm}]

\tikzstyle{brace}=[-, decorate, decoration={brace, amplitude=10pt, mirror}, draw={rgb,255: red,97; green,97; blue,97}]
\tikzstyle{arrow}=[<-, draw=black, >=latex]
\tikzstyle{dotted-line}=[-, style= dotted]
\tikzstyle{dashed-line}=[-, style=dashed]

\pgfsetlayers{nodelayer,edgelayer}
\newcommand*{\region}[1]{%
   \vspace{-1.4mm}{\color{darkgray}\hrule%
   \emph{#1}\vspace{-0.3mm}}%
}
\newcommand*{\spaceyregion}[1]{%
   \vspace{0.5mm}{\color{darkgray}\hrule%
   \vspace{1mm}\emph{#1}\vspace{-0.5mm}}%
}
\newcommand*{\regionAuxiliary}{\region{Auxiliary definitions}}
\newcommand*{\regionSkeleton}{\region{Inference skeleton}}
\newcommand*{\regionOperations}{\region{Inference operations}}
\newcommand*{\regionModelHandler}{\region{Model interpreter}}
\newcommand*{\regionModelHandlerType}{\region{Model interpreter type}}
\newcommand*{\regionInferenceHandler}{\region{Inference handler}}
\newcommand*{\regionConcreteAlgorithm}{\region{Concrete algorithm}}
\newcommand*{\inferenceTitle}[1]{\hfill{\sc#1}\vspace{-3.3mm}}
\newcommand*{\patternTitle}[1]{\inferenceTitle{Inference Pattern: #1}}
\newcommand*{\instanceTitle}[1]{\inferenceTitle{Pattern Instance: #1}}

\makeatletter
\newcommand{\thickbar}{\mathpalette\@thickbar}
\newcommand{\@thickbar}[2]{{#1\mkern1.5mu\vbox{
  \sbox\z@{$#1\mkern-1.5mu#2\mkern-1.5mu$}%
  \sbox\tw@{$#1\overline{#2}$}%
  \dimen@=\dimexpr\ht\tw@-\ht\z@-.8\p@\relax
  \hrule\@height.8\p@ 
  \vskip\dimen@
  \box\z@}\mkern1.5mu}
}
\makeatother

\newcommand{\ProbFX}{ProbFX\xspace}
\newcommand{\OurLibrary}{Our library\xspace}

\newcommand{\secref}[1]{\autoref{sec:#1}}

\newcommand{\figref}[1]{\autoref{fig:#1}}

\newcommand{\hstype}[1]{{\color{haskelltype}\textsf{#1}}}
\newcommand{\hskw}[1]{{\color{haskellkw}\textsf{#1}}}
\definecolor{bazaar}{rgb}{0.6, 0.47, 0.48}
\definecolor{string}{HTML}{B06500}
\definecolor{britishracinggreen}{rgb}{0.0, 0.26, 0.15}
\definecolor{typeconstr}{HTML}{4713A9}
\definecolor{obs}{HTML}{23424D}

\definecolor{modelCol}{HTML}{d3e5f5}
\definecolor{inferenceCol}{HTML}{8ec3de}

\newcommand{\Prob}{\mathbb{P}}

\makeatletter
\newcommand*\bigcdot{\mathpalette\bigcdot@{.6}}
\newcommand*\bigcdot@[2]{\mathbin{\vcenter{\hbox{\scalebox{#2}{$\m@th#1\bullet$}}}}}
\makeatother

\makeatletter
\newcommand*\medcdot{\mathpalette\bigcdot@{.4}}
\newcommand*\medcdot@[2]{\mathbin{\vspace*{\fill}{\hbox{\scalebox{#2}{$\m@th#1\bullet$}}}}}
\makeatother
\newcommand*\code[1]{{\small\textsf{#1}}}

\newcounter{nalg}[section] 
\renewcommand{\thenalg}{\thesection .\arabic{nalg}} 
\DeclareCaptionLabelFormat{algocaption}{Algorithm \thenalg} 

\newcommand{\tikzmark}[1]{\tikz[overlay,remember picture] \node (#1) {};}
\definecolor{blue(ryb)}{rgb}{0.01, 0.28, 1.0}
\definecolor{haskelltype}{HTML}{4713A9}
\definecolor{haskellkw}{HTML}{132AA9}
\lstdefinestyle{haskell}
{
  mathescape=true,
  frame=none,
  xleftmargin=2pt,
  stepnumber=1,
  numbersep=5pt,
  numberstyle=\ttfamily\tiny\color[gray]{0.3},
  belowcaptionskip=\bigskipamount,
  captionpos=b,
  language=haskell,
  tabsize=2,
  literate=
    {->}{$\rightarrow{}$}{1}
    {<-}{$\leftarrow{}$}{1}
    {>=>}{\code{>=>} }{1}
    {'}{\texttt{'}}{1}
    {~}{$\sim$}{1}
    {;}{$\textbf{;}$}{1}
    {=>}{$\Rightarrow{}$}{1},
  emphstyle={\bf},
  commentstyle=\it,
  stringstyle=\color[HTML]{B06500}\sffamily,
  showspaces=false,
  keywordstyle=\color[HTML]{132AA9}\sffamily\bfseries,
  columns=flexible,
  basicstyle=\small\sffamily,
  showstringspaces=false,
  morecomment=[l]\%,
  deletekeywords={words, String, foldl1, length, zip, subtract, zipWithM, uncurry, elems, Read, exec, lookup, unzip, randomR, sum, delete, floor, replicate, map, insert, fst, snd, concatMap, concat, Nothing, const, Just, fromMaybe, head, fst, mapM, Int, Double, Monad, Bool, IO, log, mapAndUnzipM, random, Either, Maybe, foldl, return, Right, zipWith, union, Left, show, List, True, False, exp, filter, elem},
  classoffset=1,
  morekeywords={uncurry, fst, snd, map, foldl1, fromMaybe, concat, findWithDefault, concatMap, length, lookupWithDefault, zip, filter, findOrInsert, filterKey, filterByKey, elem, const, head, log, fmap, size, unzip, subtract, intersection, mapAndUnzipM, lookup,  exp, keys, insert, foldl, fold, sum, elems, delete, intersectionWith, empty, replicate, replicateM, mapM, zipWith, union, forM},
  keywordstyle=\color[HTML]{132AA9}\sffamily,
  classoffset=2,
  morekeywords={pattern, defquery, @model, end, function,  family, forall},
  keywordstyle=\color[HTML]{132AA9}\sffamily\bfseries,
  classoffset=3,
  morekeywords={GuidedModel, GuidedExec, Key, VecFor, FreeTSampler, FreeT, Identity, SamF, DMap, Maybe, Deterministic, Symbol, Tracing, Static, Particle, ListT, Sequential, Traced, TraceM, FreeT, Weighted, StateT, W, Tr, Coroutine, MonadSamp, MSamp, MonadCond, MCond, GradTrace, GradUpdate, Int, Double, Bool, Guides, Grad, Handler, RW, GuideExec,  LatVar, PrtTrace, DiffDistribution, DiffDist, Arity, Nat, Vec, GuidedSample, GuidedSample, PState, Type, Constraint, ObsReader, ModelEnv, ReportedInf, Population, Either, IO, EffectSum, Support, Freer, Free, Effects, Comp, Idx, FreeF, Model, TransModel, ObsModel, FreeT, Reader, AffReader, Assign, Any, OpenProduct, Vector, Var, ObsVar, Env, MRecord, Sample, Observe, State, Addr, PrimVal, SMap, SampleTrace, LPMap, TransParams, SIREnv, ObsParams, ParamsTrans, ParamsObs, Map, MHTrace, Tag, RecInf, List, Writer, Latent, Observed, Trace, LPTrace, Params, SIRenv, Normal, Log, Gamma, Propose, ProbProg, Resample, ModelExec, ModelStep, LogP, Base, Bernoulli, Reject, NonDet, Dist, Distribution, Member, Monad, FindElem, KnownSymbol, IsLabel, HasIO, LookupType, Lookup, Observable, missing, Monoid, Observables, AsMaybes, AsLists, InfEffect},
  keywordstyle=\color[HTML]{4713A9},
  classoffset=4,
  morekeywords={},
  keywordstyle=\color[HTML]{132AA9},
  classoffset=5,
  morekeywords={<:>},
  keywordstyle=\ttfamily,
  escapechar=|
}
\lstnewenvironment{hslisting}
  {
  \lstset{
    style=haskell
    }
  }
  {}
\lstnewenvironment{hslistingsmall}
  {
  \lstset{
    style=haskell,
    basicstyle=\footnotesize\sffamily
    }
  }
  {}

\lstdefinestyle{python}{
    basicstyle=\footnotesize\ttfamily,
    mathescape=true,
    keywordstyle=\footnotesize\ttfamily\bfseries,
    language=Python
}
\lstnewenvironment{pythonlisting}
  {
  \lstset{
    style=python
    }
  }
  {}

\lstdefinestyle{algorithm}
{
  mathescape=true,
  numberstyle=\small,
  basicstyle=\footnotesize\rmfamily,
  commentstyle=\ttfamily,
  morecomment=[l]{//},
  keywordstyle=\color{black}\bfseries,
  keywords={,input, output, return, datatype, function, in, if, then, for, else, let, while, begin, end, } 
  numbers=left,
  xleftmargin=.04\textwidth
}

\lstnewenvironment{algorithm}[1][] 
  {
      \refstepcounter{nalg} 
      \captionsetup{labelformat=algocaption,labelsep=colon} 
      \lstset{ 
        style=algorithm
      }
  }
  {}
\newcommand\hsinline{\lstinline[style=haskell, columns=fullflexible]}

\begin{document}

\title[Effect Handlers for Programmable Inference]{Effect Handlers for Programmable Inference}

\author{Minh Nguyen}
\email{min.nguyen@bristol.ac.uk}
\orcid{0000-0003-3845-9928}
\affiliation{%
  \institution{University of Bristol}
  \city{Bristol}
  \country{UK}
}
\author{Roly Perera}
\email{roly.perera@bristol.ac.uk }
\orcid{0000-0001-9249-9862}
\affiliation{%
  \institution{University of Bristol}
  \city{Bristol}
  \country{UK}
}
\author{Meng Wang}
\email{meng.wang@bristol.ac.uk }
\orcid{0000-0001-7780-630X}
\affiliation{%
  \institution{University of Bristol}
  \city{Bristol}
  \country{UK}
}
\author{Steven Ramsay}
\email{steven.ramsay@bristol.ac.uk}
\orcid{0000-0002-0825-8386}
\affiliation{%
  \institution{University of Bristol}
  \city{Bristol}
  \country{UK}
}

\citestyle{acmauthoryear}
\begin{abstract}
Inference algorithms for probabilistic programming are complex imperative programs with many moving parts.
Efficient inference often requires customising an algorithm to a particular probabilistic model or problem,
sometimes called \emph{inference programming}. Most inference frameworks are implemented in languages that
lack a disciplined approach to side effects, which can result in monolithic implementations where the
structure of the algorithms is obscured and inference programming is hard. Functional programming with typed
effects offers a more structured and modular foundation for programmable inference, with monad transformers
being the primary structuring mechanism explored to date.

This paper presents an alternative approach to inference programming based on algebraic effects.
Using effect signatures to specify
the key operations of the algorithms, and effect handlers to modularly interpret those operations for specific
variants, we develop two abstract algorithms, or \emph{inference patterns}, representing two important classes
of inference: Metropolis-Hastings and particle filtering. We show how our approach reveals the algorithms'
high-level structure, and makes it easy to tailor and recombine their parts into new variants. We implement
the two inference patterns as a Haskell library, and discuss the pros and cons of algebraic effects
\emph{vis-\`a-vis} monad transformers as a structuring mechanism for modular imperative algorithm design.
\end{abstract}

\maketitle

\section{Introduction}
\label{sec:background}

Probabilistic programming languages allow modellers to use programs to formulate inference problems over models. For example in \ProbFX~\cite{probfx}, a probabilistic language embedded in Haskell, a linear regression model relating input ${x}$ and output ${y}$ linearly can be expressed as:

\begin{figure}[h]
  {\small
  \vspace{-0.2cm}
  \begin{hslisting}
  linRegr :: Double -> Double -> Model (Double, Double)
  linRegr $x$ $y$ = do
    $m$ <- call (|Sample| (|Normal| 0 3))           |\tikzmark{listing-latent-start}|
    $c$  <- call$\hspace{1mm}$(|Sample| (|Normal| 0 2))             |\tikzmark{listing-latent-end}|
    |\!|call (|Observe| (|Normal| (|$m\,$|*|$\,{x}\,$|+|$\,c$|) 1) $y$)    |\tikzmark{listing-obs-start} \tikzmark{listing-obs-end}|
    pure ($m$, $c$)
 \end{hslisting}
  }
\end{figure}

\noindent The two \code{Sample} operations specify the distributions that the line's slope $m$ and intercept
$c$ are sampled from, representing our \emph{prior} beliefs about $m$ and $c$ before accounting for any data,
denoted $\Prob(m, c)$. Given an observed output $y$ for some fixed input $x$, the operation \code{Observe}
represents a conditioning side-effect, conditioning the model against the \emph{likelihood} of $y$ having been
generated (in this case) from the normal distribution with mean $m * x + c$ and standard deviation of $1$,
denoted $\Prob(y \, | \, m, c; x)$. The variables $x$ and $y$ here are observable, whereas $m$ and $c$
that relate them are latent.

\emph{Inference} over such a model is then the process of revising our estimation of its latent
variables on the basis of the observed data, obtaining a \emph{posterior} distribution. For the linear
regression example, the Bayesian update rule yields the following equation for the posterior $\Prob(m, c \, |
\, y; x)$:

\vspace{-0.3cm}
\begin{align*}
  \underbrace{\Prob(m, c \, | \, y; x)}_\text{posterior} =  \frac{\overbrace{\Prob(y \, | \, m, c; x)}^\text{likelihood} \cdot \overbrace{\Prob(m, c)}^\text{prior}}{\underbrace{\Prob(y; x)}_\text{evidence}}
\end{align*}

\noindent
Unfortunately, extracting an exact form for the posterior is rarely simple. Although the \code{Sample} and
\code{Observe} operations in \code{linRegr} determine the prior and likelihood respectively, computing the
\emph{evidence} $\Prob(y; x)$ that forms the denominator often involves complex, high-dimensional
integration~\cite{ackerman2011noncomputable}, and probabilistic languages in practice hence use approximation
algorithms such as Monte Carlo methods \cite{andrieu2003introduction} or variational inference
\cite{fox2012tutorial}. Most techniques involve treating the model \emph{generatively} --- as something from
which samples can be drawn --- and then iteratively constraining the behaviour of the model so that, over
time, those samples eventually conform to the observations.

When using the model generatively in this way, inference algorithms need to provide their own semantics for
sampling and observing. For example, Metropolis-Hastings algorithms \cite{beichl2000metropolis} execute the
target model under specific \emph{proposals}, that fix the stochastic choices made by the model on a given
run. By selectively accepting or rejecting proposals, the algorithm controls how samples are generated, and
guarantees that as more samples are produced, the distribution of values eventually converges on the desired
posterior. Pseudocode for a generic Metropolis-Hastings iteration is shown here for linear regression:

\begin{hslisting}[mathescape=false]
   do  ($m'$, $c'$)   <- propose ($m$, $c$)
       $\rho'$       $\hspace{0.6mm}$<- exec (linRegr $x$ $y$) ($m'$, $c'$)
       b      $\hspace{0.6mm}$<- accept $\rho'$ $\rho$
       pure (if b then ($m'$, $c'$) else ($m$, $c$))
\end{hslisting}
First new values \code{$m'$} and \code{$c'$} are proposed for the slope and intercept, given the values
from the previous iteration, \code{$m$} and \code{$c$}. The function \hsinline{exec} then executes the
linear regression model with a custom semantics for sampling and observing, ensuring that \code{$m'$} and
\code{$c'$} are used for the corresponding \code{Sample} operations, and conditioning with observations
\code{$x$} and \code{$y$}. The resulting likelihood \code{$\rho'$} is compared with \code{$\rho$} from
the previous iteration to determine whether to accept the new proposal or keep the current one. Running this
procedure for many iterations will generate a sequence of samples \code{$m$} and \code{$c$} that
approximate the posterior distribution $\Prob(m, c \, | \, y; x)$.

We think of Metropolis-Hastings, as sketched here, as an inference \emph{pattern} rather than an inference
\emph{algorithm}: there are many algorithmic variants with this particular structure, differing only in how
they implement \code{propose}, \code{exec}, and \code{accept}. Indeed, most algorithms come in similar
families of variants, with abstract operations and skeletal behaviour shared by the variants, as well as their
own bespoke execution semantics for models. Particle filters \cite{Djuric03}, for example, also called
sequential Monte Carlo methods, rely on being able to \emph{partially} execute collections of models called
\emph{particles} from observation point to observation point; at each observation, particles are randomly
filtered, or \emph{resampled}, to retain only those likely to have come from the target posterior. Different
instances of the Particle Filter pattern vary in how the resampling operation works, and how particles are
executed between observation points; different choices yield different well-known algorithms.

The task of implementing these algorithmic variants falls not just to library designers; model authors
also often need to be versed in the intricacies of inference to achieve acceptable performance.
Programming new inference algorithms out of reusable parts of existing ones is sometimes called
\emph{inference programming}~\cite{venture}. Existing approaches include:
Venture~\cite{Mansinghka18}, a Lisp-based language using metaprogramming techniques; MonadBayes~\cite{monadbayes}, which uses monad transformers to implement a modular library for inference programming in Haskell; and Gen~\cite{genjl}, a inference programming framework in
Julia which relies on a fixed black-box interface for executing models generatively.

\begin{figure}
   \hspace*{-0.3cm}
   \centering
   \resizebox{1.05\columnwidth}{!}{
     \centering\captionsetup[subfigure]{justification=centering}
     \begin{tikzpicture}
	\begin{pgfonlayer}{nodelayer}
		\node [style=abstract-model] (3) at (-5.5, 9.5) {\small {Model}\\
		\; \\};
		\node [style=none] (4) at (-1, 7) {};
		\node [style=rect-white-fixedheight] (32) at (-7, 9) {\code{Observe}};
		\node [style=rect-white-fixedheight] (33) at (-4, 9) {\code{Sample}};
		\node [style=abstract-inference] (34) at (-12, 4.5) {\small {Metropolis-Hastings}\\
		\; \\};
		\node [style=abstract-inference] (35) at (-1, 4.5) {\small {Particle Filter}\\
		\; \\};
		\node [style=none] (38) at (-12, 7) {};
		\node [style=rectangle] (39) at (-13.25, 4) {\code{Propose}};
		\node [style=rectangle] (40) at (-10.5, 4) {\code{Accept}};
		\node [style=rectangle] (41) at (-1, 4) {\code{Resample}};
		\node [style=inference] (43) at (-16, 0.5) {\footnotesize \baselineskip=7pt Independence\\ Metropolis \par};
		\node [style=inference] (44) at (-8, 0.5) {\footnotesize \baselineskip=7pt  Particle\\ Metropolis-Hastings \par};
		\node [style=inference] (45) at (-3, 0.5) {\footnotesize \baselineskip=7pt  Multinomial\\ Particle Filter \par};
		\node [style=inference] (49) at (-12, 0.5) {\footnotesize \baselineskip=7pt Single-Site\\ Metropolis-Hastings \par};
		\node [style=inference] (50) at (1, 0.5) {\footnotesize \baselineskip=7pt Resample-Move\\ Particle Filter \par};
		\node [style=none] (51) at (-12, 2) {};
		\node [style=none] (52) at (-16, 2) {};
		\node [style=none] (53) at (-8, 2) {};
		\node [style=none] (62) at (-4, 7.5) {\small \emph{interprets}};
		\node [style=none] (64) at (-10.5, 2.5) {\small \emph{interprets}};
		\node [style=square-abstract-model] (76) at (-18.75, 10.5) {};
		\node [style=square-abstract-inference] (77) at (-18.75, 9.5) {};
		\node [style=square-inference] (78) at (-18.75, 7.5) {};
		\node [style=rect-dummy-fixed] (79) at (-12, 10.5) {\small Abstract modeling code};
		\node [style=rect-dummy-fixed] (80) at (-12, 9.5) {\small Abstract inference pattern};
		\node [style=rect-dummy-fixed] (81) at (-12, 7.5) {\small Inference algorithm};
		\node [style=square-white] (82) at (-18.75, 8.5) {};
		\node [style=rect-dummy-fixed] (83) at (-12, 8.5) {\small Operation};
		\node [style=none] (85) at (-5.5, 7) {};
		\node [style=none] (88) at (-1, 2) {};
		\node [style=none] (89) at (-3, 2) {};
		\node [style=none] (90) at (1, 2) {};
		\node [style=none] (91) at (0.5, 2.5) {\small \emph{interprets}};
		\node [style=none] (93) at (-18.25, 6.25) {};
		\node [style=none] (94) at (-5.75, 6.25) {};
		\node [style=none] (95) at (-5.75, -1.25) {};
		\node [style=none] (96) at (-18.25, -1.25) {};
		\node [style=none] (97) at (-12, 6.25) {};
		\node [style=none] (98) at (-5, 6.25) {};
		\node [style=none] (99) at (-5, -1.25) {};
		\node [style=none] (100) at (3.25, -1.25) {};
		\node [style=none] (101) at (3.25, 6.25) {};
		\node [style=none] (102) at (-1, 6.25) {};
	\end{pgfonlayer}
	\begin{pgfonlayer}{edgelayer}
		\draw (52.center) to (51.center);
		\draw (53.center) to (51.center);
		\draw [style=arrow] (34) to (51.center);
		\draw (52.center) to (43);
		\draw (53.center) to (44);
		\draw [style=arrow] (35) to (88.center);
		\draw (89.center) to (88.center);
		\draw (88.center) to (90.center);
		\draw (45) to (89.center);
		\draw (50) to (90.center);
		\draw (49) to (51.center);
		\draw [style=dotted-line] (93.center) to (96.center);
		\draw [style=dotted-line] (96.center) to (95.center);
		\draw [style=dotted-line] (94.center) to (95.center);
		\draw [style=dotted-line] (93.center) to (97.center);
		\draw [style=dotted-line] (97.center) to (94.center);
		\draw (97.center) to (38.center);
		\draw [style=dotted-line] (98.center) to (99.center);
		\draw [style=dotted-line] (99.center) to (100.center);
		\draw [style=dotted-line] (100.center) to (101.center);
		\draw [style=dotted-line] (98.center) to (102.center);
		\draw [style=dotted-line] (102.center) to (101.center);
		\draw (102.center) to (4.center);
		\draw (38.center) to (85.center);
		\draw (85.center) to (4.center);
		\draw [style=arrow] (3) to (85.center);
	\end{pgfonlayer}
\end{tikzpicture}
   }
   \vspace{-0.2cm}
   \caption{Inference patterns presented in this paper}
   \label{fig:inference-patterns}
   \vspace{-0.5cm}
\end{figure}

In this paper, we present an approach to programmable inference based on algebraic effects.
We use effect signatures to specify the key operations of various classes of abstract inference algorithms,
and effect handlers to specialise those algorithms into concrete variants adapted to specific problems.
We use the approach to develop two abstract algorithms, or \emph{inference patterns}, representing two
important classes of inference, and implement them in Haskell. Our specific contributions are as follows:

\begin{itemize}[leftmargin=5mm]
   \item \secref{inference-patterns} informally introduces the idea of an inference pattern.
   \item \secref{metropolis} presents the \emph{Metropolis-Hastings} inference pattern, with Independence Metropolis and Single-Site Metropolis-Hastings as illustrative instances.
   \item \secref{particle} presents the \emph{Particle Filter} inference pattern, also known as sequential
   Monte Carlo, with Multinomial Particle Filter and Resample-Move Particle Filter as instances. We also
   derive Particle Metropolis-Hastings, an instance of Metropolis-Hastings which uses Particle Filter.
   \item \secref{performance} shows that the performance of our approach is competitive with
   state-of-the-art systems for programmable inference based on other techniques.
   \item \secref{eval-modularity} 
   contrasts our approach to untyped approaches such as Gen and Venture, and MonadBayes, the main
   existing framework based on typed effects.
\end{itemize}

The algorithms we discuss are well known; what we bring to the picture is the novel modular architecture,
outlined in \figref{inference-patterns}, which reveals the high-level structure of the algorithms and makes it
easy to tailor and recombine their parts into new variants. The \emph{model}, provided by the user, expresses
an abstract inference problem in terms of \code{Sample} and \code{Observe} operations. \emph{Inference
patterns}, provided by library designers, assign specific semantics to those operations, and provide skeletal
procedures for iteratively executing a model under those semantics. These procedures are in turn expressed in
terms of their own abstract operations, which can also be assigned a semantics to obtain a concrete algorithm
capable of generating samples from the model's posterior.

This design makes it easy to define new algorithmic variants out of existing ones. For example, we can easily
build a particle filter algorithm which uses another well-known inference algorithm, Metropolis-Hastings, as
an internal component; equally easily, we can derive a version of Metropolis-Hastings that uses particle
filter. Moreover each of these complex scenarios arise in real-world solutions.


\hspace{-0.1cm} We build on two pieces of prior work: the \emph{extensible freer monad}~(\secref{background:extensible-effects}), which
adds an extensible effect system to Haskell, and
\ProbFX (\secref{background:effects-for-models}), an embedding of probabilistic models in Haskell based on this~approach.

\subsection{\hspace{-0.2cm} Background: An Embedding of Extensible Effects}
\label{sec:background:extensible-effects}

Effect systems model effects as coroutine-like interactions between side-effecting expressions that request
\emph{operations} to be performed, and special contexts, called \emph{handlers}, that assign meaning to those
operations~\cite{bauer2015programming}. An operation may provide a continuation, allowing the handler to return
control to the requesting expression. Effect systems offer a flexible alternative to monad
transformers~\cite{Liang95} for adding complex imperative features to functional~languages, making them an
appealing tool for structuring inference algorithms. But the only precedent we know of
is by \citet{scibior2015effects} on basic rejection~sampling.

\begin{figure}
  \input{code/background/comp}
  \vspace{-0.2cm}
  \caption{Extensible freer monad embedding}
  \label{fig:comp}
  \vspace{-0.3cm}
  \end{figure}
The \emph{extensible freer monad}~\cite{kiselyov2015freer} is an embedding of a typed effect system into
Haskell, exploiting Haskell's rich support for embedded languages. The basic idea is to represent an effectful
computation using the recursive datatype \hsinline{Comp es a} at the top of \figref{comp}. A term
of type \hsinline{Comp es a} represents a computation that produces a value of type \code{a}, whilst
possibly performing any of the computational effects specified by the \emph{effect signature} \code{es}, a
type-level list of type constructors. Leaf nodes \code{Val x} contain pure values \code{x} of type
\code{a}. Operation nodes \code{Op op k} contain \emph{operations} \code{op} of the abstract datatype
\hsinline{EffectSum es b}, representing the invocation of an operation of type \code{e b} for some effect
type constructor \code{e} in \code{es}, where \code{b} is the (existentially quantified) return type of
the operation; the argument \code{k} is a continuation of type \code{b} $\rightarrow$ \hsinline{Comp es a}
that takes the result of the operation and constructs the remainder of the computation.

As one might surmise, \hsinline{Comp es} is a monad, allowing effectful code to piggyback on Haskell's
\hsinline{do} notation for sequential chaining of monadic computations. The bind operator \hsinline{(>>=)} can
be viewed as taking a computation tree of type \hsinline{Comp es a} and extending it at its leaves with a
computation generated by \code{f :: }\hsinline{a -> Comp es b}. In the \code{Val x} case, a new
computation \code{f x} is returned; otherwise for \code{Op op k}, the rest of the computation \code{k}
is composed with \code{f} using Kleisli composition \code{(>=>)}. Values of type \hsinline{Comp es a}
are thus uninterpreted ``computation trees'' comprised of pure values and operation calls chosen from
\code{es}.

\hsinline{EffectSum es} is key to the extensibility of the approach, representing an ``open'' (extensible) sum
of effect type constructors; a concrete value of type \hsinline{EffectSum es a} is an operation of type
\code{e a} for exactly one effect type constructor \code{e} contained in \code{es}.
The implementation of \hsinline{EffectSum} is hidden; the type class \code{$\in$} provides methods for
safely injecting and projecting effectful operations of type \code{e a} into and out of \hsinline{EffectSum es a}, with the constraint \hsinline{e $\in$ es} asserting that \code{e} is a member of \code{es}. The
helper \code{call} makes it easy to write imperative code, as we saw in the \code{linRegr} example,
injecting the supplied operation into \hsinline{EffectSum es a} and supplying the leaf continuation
\hsinline{Val}.

\subsubsection{Interpreting Effectful Computations}

\begin{figure}
\input{code/background/handle-comp}
\vspace{-0.3cm}
\caption{Effect handlers and \code{handle}/\code{handleWith} helpers}
\label{fig:handle}
\vspace{-0.4cm}
\end{figure}

\figref{comp} provided the machinery required to construct effectful computations; \figref{handle} shows the
machinery required to execute them. Executing an effectful computation means providing a ``semantics'' for
each of its effects, in the form of an interpreter called an \emph{effect handler}. A handler for effect type
\code{e} has the type \hsinline{Handler e es a b}; it assigns partial meaning to a computation tree by
interpreting all operations of type \code{e}, \emph{discharging} \code{e} from the front of the effect
signature, and transforming the result type from \code{a} to \code{b}. Effect handlers are thus modular
building blocks which compose to constitute full interpretations of programs.

The helpers \code{handle} and \code{handleWith} make it easy to implement handlers; \code{handleWith} is
used for handlers that also thread a state of type \code{s}, whereas \code{handle} sets \code{s} to be
the trivial unit type. Both take two higher-order arguments: \hsinline{hval}, which says how to interpret pure
values, and \hsinline{hop}, which says how to interpret operations of effect type \code{e}. In the
\code{Val x} case, where the computation contains no operations, we simply apply \hsinline{hval} to the
return value (and state), yielding a computation from which \code{e} has been discharged. In the \code{Op
op k} case, where \code{op} has type \hsinline{EffectSum (e : es) a}, the auxiliary function \code{decomp}
determines whether \code{op} belongs to the leftmost effect \hsinline{e}, and can thus be handled by
\code{hop}, or whether it belongs to an effect in \code{es}, in which case we can simply reconstruct the
operation at the narrower type. In either case we recurse (by extending the continuation) to ensure that the
rest of the computation is handled similarly.

\subsection{Effects for Probabilistic Models}
\label{sec:background:effects-for-models}

\citet{probfx} use the extensible freer monad representation from \secref{background:extensible-effects} to
define an embedding of probabilistic models. \emph{Models}, in \figref{model-definition}, are simply
computations with access to two specific effects, \hsinline{Sample} and \hsinline{Observe}, each with one
operation: \code{Sample d} samples from probability distribution \code{d}, and \code{Observe d y} conditions
\code{d} on an observed value \code{y} before returning that same value \code{y}. These operations
characterise the minimal interface assumed by most inference methods, and for simplicity here, we assume they
are the \emph{only} model effects required, along with \hsinline{IO} for random number~generation.

\begin{figure}
\input{code/background/model}
\vspace{-0.3cm}
\caption{Models as computations that sample and observe}
\label{fig:model-definition}
\vspace{-0.2cm}
\end{figure}

Both \code{Sample} and \code{Observe} are constrained by the type class \hsinline{Dist d a},
specifying that the type \code{d} represents a primitive distribution generating values of type \code{a},
with the functional dependency \hsinline{d -> a} indicating that \hsinline{d} fully determines \hsinline{a}.
Instances of \hsinline{Dist d a} must implement two functions: (i) \code{draw}, which takes a distribution
\code{d} and random point \code{r} from the unit interval $[0,1]$, and draws a sample by inverting
the cumulative distribution function of \code{d} at \code{r}; and (ii) \code{logProb}, which
computes the log probability of \code{d} generating a particular value. (The synonym \hsinline{LogP} is
helpful for distinguishing log probabilities from other values of type \hsinline{Double}.) For example, the Bernoulli distribution over Booleans, with probability \code{$p$} of generating \code{True}, and
\code{1 - $p$} for \code{False}, can be implemented as:
\begin{lstlisting}[style=haskell, escapechar=*]
data Bernoulli = *Bernoulli* { $p$ :: Double }
instance Distribution Bernoulli Bool where
  draw    *(Bernoulli $p$)* r = r $\leq$ $p$
  logProb *(Bernoulli $p$)* b = if b then log $p$ else log (1 - $p$)
\end{lstlisting}

\noindent
This states that \code{draw}ing \code{True} corresponds to drawing a random value \code{r $\leq$ $p$}
uniformly from \code{[0, 1]}, with the log probabilities \code{log $p$}  and \code{log (1 - $p$)} of
drawing \code{True} and \code{False} respectively.

\subsubsection{Interpreting Probabilistic Models}
\label{sec:handling-models}

Interpreting a model means providing a semantics for \hsinline{Sample} and \hsinline{Observe}. The most basic
interpretation of a model, as a generative process with no inference, is usually called \emph{simulating} (or
\emph{sampling from}) the model, and can be defined as the composition of the handlers shown in
\figref{simulate}. The handler \code{defaultObserve} (trivially) interprets \code{Observe d y} operations
to return the observed value \code{y}, via the continuation. The handler \code{defaultSample} interprets
\code{Sample d} operations, as long as \hsinline{IO} is also
present in the effect signature, by first drawing a random value \code{r} uniformly from the interval
\code{[0, 1]} using the \hsinline{IO} function \hsinline{random}, and then generating a sample from
\code{d} using \code{draw}. Lastly, the function \code{runIO} discharges the final \hsinline{IO} effect by
simply extracting and sequencing the \hsinline{IO} actions, running the computation as a top-level Haskell program.

\begin{figure}
\input{code/background/handle-model}
\vspace{-0.35cm}
\caption{Effect handlers for model simulation}
\label{fig:simulate}
\vspace{-0.3cm}
\end{figure}

\section{Inference Patterns}
\label{sec:inference-patterns}

\begin{figure*}
\input{figures/background/inference-pattern}
\vspace{-0.4cm}
\caption{Inference patterns (left) and pattern instances (right)}
\label{fig:inference-pattern}
\vspace{-0.25cm}
\end{figure*}

Our approach to programmable inference builds on the general embedding of extensible effects from
\secref{background:extensible-effects} and  probabilistic models from
\secref{background:effects-for-models}. Our key insight is that algebraic effects seem to be a natural fit for
two kinds of extensibility central to programmable inference. First, representing models as (reinterpretable)
effectful computations allows them to be assigned semantics tailored to specific algorithms. For example, we
can instrument models to produce the traces needed for Metropolis-Hastings (\secref{metropolis}), or arrange
for models to execute stepwise rather than to completion for particle filters (\secref{particle}). Second, we
can take a similar view of the algorithms themselves. By representing the key actions of each broad approach
to inference as reinterpretable ``inference operations'' --- for example resampling, in the case of particle
filters --- we can turn them into extension points that can be given different meanings by different members
of the same broad family of algorithms. Deriving a concrete inference algorithm is then a matter of supplying
appropriate interpreters for the model and for the inference operations themselves. Moreover these extension
points advertise to non-experts the key steps in the algorithms.

As well as offering a modular and programmable approach to algorithm design, this perspective also provides a
useful conceptual framework for understanding inference. For example, Metropolis-Hastings and particle filters
might look quite different algorithmically, but our approach provides a uniform way of looking at them: each
can be understood as an abstract algorithm, parameterised by a model interpreter, and expressed using abstract
operations whose interpretation is deferred to concrete implementations. This informal organisational
structure we call an \emph{inference pattern}, and is shown on the left-hand side of
\figref{inference-pattern}; a library designer developing their own abstract inference algorithms using our
approach would most likely follow this high-level template. We now flesh out the idea of an inference pattern
a little before turning to the patterns we developed for this paper.

\paragraph{Inference patterns}

The core of an inference pattern (see \figref{inference-pattern}, left) is an abstract algorithm expressing an
inference procedure. Taking inspiration from the parallelism literature \cite{Darlington95}, we call this an
\emph{inference skeleton}. Inference skeletons depend on algebraic effects in two essential ways. First, each
skeleton is parameterised by a \emph{model interpreter}, giving concrete algorithms control over model
execution; second, the skeleton is expressed in terms of abstract \emph{inference operations} unique to the
pattern, which act as additional extension points where concrete algorithms can plug in specific behaviour.

The model interpreter has a \emph{model interpreter type}, whose exact form depends on the pattern, but is
roughly:
\begin{lstlisting}[style=haskell]
  type ModelExec a b = Model a -> IO b
\end{lstlisting}
\noindent and is used by the skeleton to fully interpret the model into an \hsinline{IO} action on each
iteration. 
Having the inference skeleton execute the model all the way to an \hsinline{IO} action allows the model and
inference algorithm to have distinct effect signatures. Assuming inference operations with concrete type
\hsinline{InfEffect}, a skeleton will have a type resembling:

\begin{lstlisting}[style=haskell, escapechar=*]
  *infSkeleton \,::\, (\hstype{InfEffect} $\in$ fs, \hstype{IO} $\in$ fs)*
    => ModelExec a b -> Model a -> Comp fs b
\end{lstlisting}

\noindent where \code{fs} contains only the effects specific to the algorithm. If instead the skeleton were
to incorporate the effects of the model into its own computation, and the interpretation of the model deferred
until the handling of the inference operations, \code{fs} would need to include model operations like
\hsinline{Observe} and \hsinline{Sample}, and the resulting computation trees would be much larger.
Keeping the effect signatures distinct makes for a more modular and efficient design.

\paragraph{Pattern instances}

A pattern instance (\figref{inference-pattern}, right) provides a \emph{concrete algorithm}. It instantiates
an inference skeleton with a specific model interpreter, determining the specific model execution semantics to
be used, and then composes the result with an \emph{inference handler} providing a specific interpretation of
the inference operations. Pattern instances may also have auxiliary definitions.

\vspace{3mm} We use this informal template to present our two inference patterns:
Metropolis-Hastings~(\secref{metropolis}) and Particle Filter~(\secref{particle}), along with concrete
instances illustrating the compositionality and programmability of the approach. These are available
in an open source Haskell~library.~\footnote{\href{https://github.com/min-nguyen/prob-fx-2}{github.com/min-nguyen/prob-fx-2}}

\vspace{-0.05cm}

\section{Inference Pattern: Metropolis-Hastings}
\label{sec:metropolis}

Metropolis-Hastings algorithms repeatedly draw samples from a
chosen ``proposal'' distribution. How these samples are generated is controlled by an \mbox{accept/reject} scheme, determining whether to accept a
new proposal and thus move to a new configuration, or to reject it and remain in the current configuration. Under certain standard assumptions, then, these samples yield a Markov
chain that converges to the target posterior. (Here we only consider the case where the proposal distribution
is the actual model we are performing inference over.)

The key operations of the algorithm are proposing and accepting/rejecting proposals. To expose them as
extension points, we represent them by the inference effect  \hsinline{Propose w} in \figref{metropolis}.
The parameter \code{w} is a particular representation of probability, or \emph{weight}; the datatype \hsinline{Trace}
represents proposals. A trace fixes a subset of the stochastic choices made by a model, which is key to how
the algorithm controls where samples are drawn from.

\begin{figure}
   \input{code/metropolis/pattern}
   \vspace{-0.35cm}
\caption{Inference Pattern: Metropolis-Hastings}
\label{fig:metropolis}
\vspace{-0.3cm}
\end{figure}

The inference skeleton \code{mh n $\tau_0$} executes \code{n} abstract iterations of Metropolis-Hastings,
iterating \code{mhStep} to generate a Markov chain of length \code{n}, from a (typically empty) starting
trace \code{$\tau_0$}. The head of the Markov chain \code{(x, ($w$, $\tau$))} represents the current configuration;
\code{x} is the sample last drawn from the model, \code{$\tau$} is the trace for that model run, and
\code{$w$} is an associated weight of type \code{w}, representing the probability density at
\code{$\tau$}. First, \code{mhStep} calls \code{Propose $\tau$} to generate a new proposal
\code{$\tau^\dagger$} derived from \code{$\tau$}. Then, the model interpreter \code{exec} is used to run the
\hsinline{Model} (\figref{model-definition}), using the information in \code{$\tau^\dagger$} to fix stochastic choices, and resulting in a new trace
\code{$\tau'$} and associated weight \code{$w'$}. The new trace contains at least as much information as
\code{$\tau^\dagger$}, but additionally
stores any choices not determined by \code{$\tau^\dagger$}. The result of \code{exec} is an \hsinline{IO}
computation, which is inserted into the computation tree using \hsinline{call}. Finally, \code{mhStep}
calls \code{Accept} to determine whether the new configuration is by some (unspecified) measure ``better'' than
the current one, returning it if so, and otherwise retaining the current.

To fix stochastic choices, a trace must associate to each \code{Sample} operation enough information to
determinise that sample. This can be achieved in various ways, but here we assume that \code{Sample} nodes
are identified by \emph{addresses} \code{$\alpha$} \cite{anglican} of abstract type \hsinline{Addr},
either generated behind the scenes or manually assigned by the user;
a trace is then a map from addresses to random values \code{r $\in$ [0, 1]} providing the source
of randomness for drawing the sample associated with a given address.
The \hsinline{Sample} handler \code{reuseTrace $\tau$} is used for executing a model under a trace \hsinline{$\tau$}: it generates the draw using the stored
random value for \code{$\alpha$} if there is one, and otherwise generates a fresh value \code{r} which is recorded
in an updated trace. Since \code{draw} is pure, executing a model under a fixed (and sufficiently large)
trace is deterministic, allowing the generative behaviour of the model to be controlled by providing it a
specific trace. The \code{reuseTrace} handler is thus a reusable ``inference component'' which can be used
by concrete instances of Metropolis-Hastings, of which we now present two examples: Independence Metropolis
(\secref{metropolis:im}) and Single-Site Metropolis-Hastings (\secref{metropolis:ssmh}).

\subsection{Pattern Instance: Independence Metropolis}
\label{sec:metropolis:im}

\figref{im} defines a simple Metropolis-Hastings variant called Independence Metropolis, where each iteration
proposes an entirely new set of samples, and determines whether the proposal is accepted by comparing its
likelihood with the previous iteration. This specialises the weight type \code{w} in \hsinline{Propose w}
and \hsinline{ModelExec w a} to the type \hsinline{LogP} for log likelihoods.

\begin{figure}
   \input{code/metropolis/im}
   \vspace{-0.3cm}
   \caption{Pattern Instance: Independence Metropolis}
   \label{fig:im}
   \vspace{-0.3cm}
\end{figure}

The handler \code{handlePropose$_\text{im}$} interprets \code{Propose} by mapping new random values over the entire trace. (One can equivalently return the \hsinline{empty} trace, but our particular approach becomes useful for Particle Metropolis-Hastings in \secref{particle:pmh}.)
To interpret \code{Accept}, we compute the likelihood ratio between the current and previous
iteration, and accept only if greater than a random point in the interval $[0, 1]$.

For model execution, \code{likelihood}  handles \hsinline{Observe} by summing the log likelihood \hsinline{$w$} over all observations with
\hsinline{0} as the starting value. The full Independence Metropolis algorithm is then derivable by providing
\code{mh} with a number of iterations \code{n}, the \hsinline{empty} map as the initial trace, and the model interpreter,
before post-composing with \code{handlePropose$_\text{im}$} and \code{runIO} to yield a Markov
chain of \code{n} proposals for a given~model.

\subsection{\hspace{-0.2cm} Pattern~Instance:~Single-Site~Metropolis-Hastings}
\label{sec:metropolis:ssmh}

The rate of accepted proposals in Independence Metropolis suffers as more variables are sampled from: because
each proposal generates an entirely new trace, achieving a high likelihood means sampling an entire set of
likely proposals. \figref{ssmh} defines Single-Site Metropolis-Hastings \cite{wingate2011lightweight}, which
uses an alternative semantics for model execution and inference: proposing just one sample per iteration, and
otherwise reusing samples from the previous iteration. The acceptance/rejection scheme is also slightly
different, comparing individual probabilities of \code{Sample} and \code{Observe} operations with respect
to the proposed sample. This specialises the weight type \code{w} of \hsinline{Propose w} to
\hsinline{LPTrace}, mapping addresses to their log probabilities.

\begin{figure}
   \input{code/metropolis/ssmh}
   \vspace{-0.3cm}
   \caption{Pattern Instance: Single-Site Metropolis-Hastings}
   \label{fig:ssmh}
   \vspace{-0.35cm}
\end{figure}

The handler \code{handlePropose$_\text{ssmh}$} threads an address \code{$\alpha$}, identifying
the sample currently being proposed. (The initial value of this argument is unused, so we supply an arbitrary
value \code{$\alpha_0$}.) For \code{Propose}, we use a helper \code{randomFrom} to select a new address
\code{$\alpha$} uniformly from the keys of trace \hsinline{$\tau$}, and then return \code{$\tau$} updated
with a new random value for \code{$\alpha$}. For \code{Accept}, the acceptance ratio between \code{$w'$}
and \code{$w$} is computed for corresponding addresses by \hsinline{intersectionWith (-)}, using
\hsinline{delete $\alpha$} to exclude the current proposal site, and also accounting for the ratio of sizes between the two traces.
~\footnote{By using {\footnotesize\hskw{intersectionWith} (-)}, we assume that each execution of the model  encounters the same (addresses of) {\textsf{\footnotesize{Observe}}} operations, which is a common assumption in probabilistic programming languages.}
If the new trace \code{$\tau'$} is accepted then \hsinline{intersection $\tau'$ $w'$} clears all unused samples from it, given that $w'$ will only ever store addresses relevant to the model's execution, as described next.

The semantics for model execution differs only slightly from Independence Metropolis. Instead of summing the
log probabilities of \code{Observe} operations, we record the log
probabilities of all \code{Observe} and \code{Sample} operations encountered into a fresh map \hsinline{$w$} of type
\hsinline{LPTrace}, via the handler \code{traceLP} shown in \figref{ssmh}. 
This deviates from the normal handler pattern, matching on the result of \code{prj op} (\secref{background})
to intercept operations of different effect types but leaving them unhandled. Here we simply
modify the continuation \code{k} to store the log probability of the operation's result. The
case of \code{prj} returning \code{Nothing} follows the same pattern as \code{decomp} returning \code{Left} in \figref{handle}.

All conditioning side-effects are in fact taken care of by \code{traceLP}, so the residual
\hsinline{Observe} operations are handled by \code{defaultObserve} to simply return the observed
values, and the \hsinline{Sample} operations by \code{reuseTrace} as before; the interpreted model has
type \hsinline{IO (a, (LPTrace, Trace))}, containing the final log probability and execution traces.
We now have the parts to derive Single-Site Metropolis-Hastings from the \code{mh} pattern.

\vspace{-0.3cm}
\section{Inference Pattern: Particle Filter}
\label{sec:particle}

Particle filters \cite{doucet2009tutorial} generate samples from the posterior by considering \emph{partial}
model executions. The idea is to spawn multiple instances of the model called \emph{particles}, and then
repeatedly switch between (i) running the various particles in parallel up to their next observation, and (ii)
subjecting them to a \emph{resampling} process \cite{hol2006resampling}. Resampling is a stochastic strategy
for filtering out particles whose observations are deemed unlikely to have come from the posterior, i.e.~are
weighted lower than other particles. Ideally, after many resampling steps, only particles that closely
approximate the posterior will remain.

A particle filter configuration is a list of (particle, weight) pairs of type \hsinline{(Model a, w)}. The key
operation is resampling, which transforms a configuration by discarding some particles and duplicating others,
but usually keeping the number of particles constant; we expose this as an extension point via the inference
effect type \hsinline{Resample w} in \figref{particle-filter}. The model interpreter type \hsinline{ModelStep w a} for particle filter is distinctive because it characterises \emph{particle steppers}, which partially
execute particles: a particle stepper resumes a suspended particle with weight \code{w}, executes it by some
unspecified amount, and returns an updated particle and weight.

\begin{figure}
  \input{code/particle-filter/pattern}
  \vspace{-0.3cm}
  \caption{Inference Pattern: Particle Filter}
  \label{fig:particle-filter}
  \vspace{-0.3cm}
\end{figure}

The inference skeleton \code{pfilter n $w_0$} describes a generic particle filter, recursively running a set of \code{n} particles with a starting
weight of \code{$w_0$} until termination using \code{pfStep}, at each iteration using the particle stepper
\code{step} to obtain a new configuration \code{p$w$s$'$}. The function \code{done} examines the new
configuration to determine whether all particles have terminated, in which case the return values and final
weights of the particles \code{rs} are returned, or whether some particles are still executing, in which
case the algorithm calls \code{Resample} on the configuration and continues with the filtered result.

The handler \code{advance} is a reusable inference component for implementing particle steppers.
Given an initial weight \code{$w$}, it advances a particle to the next \code{Observe}, returning the rest
of the computation \code{k y} \emph{unhandled}, along with the accumulated weight at that point. Matching on
\code{Val} instead means the particle has terminated, and so is returned alongside its final weight. Notice
that \code{advance} is not implemented in terms of \code{handle}; this is because \code{handle} produces
a ``deep'' handler~\cite{hillerstrom2018shallow} which discharges the handled effect from the effect
signature, and so does not support the shallow (partial) handling needed for suspensions.

We now present two  instances of Particle Filter: Multinomial Particle Filter
(\secref{multinomial}) and Resample-Move Particle Filter (\secref{particle:rmpf}), the latter constructed
using Metropolis-Hastings. We also present Particle Metropolis-Hastings (\secref{particle:pmh}), an instance of Metropolis-Hastings
constructed using Particle Filter.

\subsection{Pattern Instance: Multinomial Particle Filter}
\label{sec:multinomial}

Many basic variants of particle filters can be implemented by recording just the log probabilities of particles,
specialising \hsinline{w} in \hsinline{Resample w} and \hsinline{ModelStep w} to \hsinline{LogP}. A popular
example is a particle filter that uses a ``multinomial resampling'' algorithm, defined in
\figref{particle-filter:mul}. To interpret \code{Resample p$w$s}, containing \code{$n$} particles
\code{ps} and their weights \code{$w$s}, we use the categorical distribution to draw \code{$n$} integers
from the range \hsinline{$0$ ... $n - 1$} with log probabilities corresponding to the normalised weights \code{$w$s$_\code{norm}$}.
These integers indicate the positions of particles to continue
executing with, which are extracted by indexing with \code{(!!)}, and then uniformly paired with the log mean of the weights, \code{$\thickbar{w\code{s}}$}; it is expected for particles with higher weight to be selected more than once, and unlikely
ones pruned.

\begin{figure}
  \input{code/particle-filter/mul-pf}
  \vspace{-0.3cm}
  \caption{Pattern Instance: Multinomial Particle Filter}
  \label{fig:particle-filter:mul}
  \vspace{-0.3cm}
\end{figure}

For model execution, \hsinline{Observe} is handled by \code{advance}, and \hsinline{Sample} simply by \code{defaultSample} for drawing random values. Then we can derive
\code{mulpfilter} by using \code{pfilter n 0 handleParticle} to construct an abstract particle filter of
\code{n} particles with starting weight \code{0}, and composing with \code{(runIO $\circ$ handleResample$_\text{mul}$)} to specialise to a multinomial particle filter that generates \code{n}
samples from the posterior and their final weights.

\begin{figure}
  \input{code/particle-filter/rmpf}
  \vspace{-0.2cm}
  \caption{Pattern Instance: Resample-Move Particle Filter}
  \label{fig:particle-filter:rmpf}
  \vspace{-0.3cm}
\end{figure}

\subsection{Pattern Instance: Resample-Move Particle Filter}
\label{sec:particle:rmpf}

Complex inference problems often require the programmer to combine different top-level inference procedures,
each addressing a different sub-problem. For example, the resampling step in particle filtering can result
in many particles becoming the same, limiting the range of values sampled from the posterior, a problem called
\emph{particle degeneracy}. One solution is to use Metropolis-Hastings proposals to ``move around'' the
sampled values of each particle after resampling, an approach called the Resample-Move Particle Filter
\cite{gilks2001following}. This kind of wholesale algorithm reuse is also supported in our framework, and we
show this now by deriving Resample-Move Particle Filter (\figref{particle-filter:rmpf}) as an instance of Particle Filter,
by providing a \hsinline{Resample} handler which calls Single-Site Metropolis-Hastings
(\secref{metropolis:ssmh}).

To specialise \code{pfilter} to use Metropolis-Hastings, we set the weight parameter \code{w} to
\hsinline{PState}, now also storing the particle's execution trace to allow for proposals. To know
how far to execute a particle under a given proposal, the \hsinline{Resample}
handler increments a state variable \code{t} at each \code{Resample}, tracking the number of observations encountered so far in the model.

To handle \code{Resample}, we unzip the particle states into their weights \code{$w$s} and traces
\code{$\tau$s}, using \code{$w$s} to carry out multinomial resampling as in \figref{particle-filter:mul} but
for resampling a selection of traces \code{$\tau$s$_\text{res}$}. The helper \code{suspendAfter}
then produces a copy \code{model$_\text{t}$} of the model suspended after observation \code{t}, which will
let us instantiate new particles that resume at that point. We execute \code{model$_\text{t}$} under each
resampled trace in \code{$\tau$s$_\text{res}$} for a series of \code{ssmh} updates; the most recent update
is taken, from which the final moved particle \code{p$_\text{mov}$} and its trace are used.

The model interpreter is simply a particle stepper which uses \code{reuseTrace} instead of \code{defaultSample} to record/reuse the particle's trace. The concrete algorithm \code{rmpf n m}
can then be assembled from these parts, yielding a multinomial particle filter of \code{n} particles,
where each resampling step is followed by \code{m} Single-Site Metropolis-Hastings updates to each particle.

\begin{figure}
  \input{code/particle-filter/pmh}
  \vspace{-0.2cm}
  \caption{Pattern Instance: Particle Metropolis-Hastings}
  \label{fig:metropolis:pmh}
  \vspace{-0.4cm}
\end{figure}

\subsection{Pattern Instance: Particle Metropolis-Hastings}
\label{sec:particle:pmh}

We now revisit the Metropolis-Hastings inference pattern from \secref{metropolis}, and show that
our framework makes it equally easy to reuse a particle filter inside Metropolis-Hastings. In \secref{metropolis}, we
only considered algorithms where the proposed traces \hsinline{$\tau$} fixed the values of \emph{all} latent
variables, fully determinising the model. But often we only care about proposing a \emph{subset} of the trace,
\hsinline{$\tau_\theta$} for some variables of interest \hsinline{$\theta$}, allowing the other latent
variables to be freshly sampled. It then becomes possible to use a particle filter to run each proposal for
many different simulations, averaging over the particles to compute the likelihood used to accept or reject
the proposal. This is known as Particle Metropolis-Hastings~\cite{dahlin2015particle} and is used to reduce
the variance of likelihood estimates of proposals. \figref{metropolis:pmh} derives a version of this from the Metropolis-Hastings pattern, providing a \hsinline{ModelExec} that also calls a multinomial particle
filter (\secref{multinomial}), and reusing the \hsinline{Propose} handler from Independence Metropolis~(\secref{metropolis:im}).

The model interpreter takes a number of particles \code{n} and trace \code{$\tau_\theta$} providing values
for the latent variables of interest, i.e.~addresses \code{$\theta$}. It begins by defining an internal
particle stepper which executes a particle to the next observation as usual, but handles \hsinline{Sample}
with \code{reuseTrace $\tau_\theta$} so that each particle uses fixed values for the latent variables in
$\theta$, using \hsinline{fmap fst} to ignore the updated trace. The particle stepper is then used to
instantiate a particle filter otherwise identical to the multinomial one, producing a list of particle outputs
\code{xs} and weights \code{$w$s}. To conform to the \hsinline{ModelExec} type for Metropolis-Hastings,
the model interpreter must return a model result plus a weight and trace; for the model result we
draw an element of \code{xs} with probability proportional to the weights, and for the weight we use the log mean of \code{$w$s}.
For the trace, we return \code{$\tau_\theta$} rather than the possibly extended trace returned by \code{reuseTrace}, to avoid fixing stochastic choices other than those in \code{$\tau_\theta$} (when handling \code{Propose} in \secref{metropolis:im}).

The algorithm \code{pmh m n $\theta$} then describes \code{m} Independence Metropolis proposals for
addresses \code{$\theta$}, but where each proposal is weighted by simulating the model as \code{n}
particles. The first two lines initialise \code{$\tau_\theta$}, using
\hsinline{reuseTrace empty} to populate an empty trace, and then filtering to the addresses in
\hsinline{$\theta$}.

\vspace{-0.075cm}
\section{Performance Evaluation}
\label{sec:performance}

\begin{figure*}
  \centering
  \begin{subfigure}{\textwidth}
    \centering
    \includegraphics[width=0.35\textwidth]{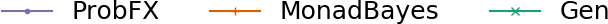}
    \vspace*{0.4cm}
  \end{subfigure}
  \begin{subfigure}{\textwidth}
    \centering
    \includegraphics[width=\textwidth]{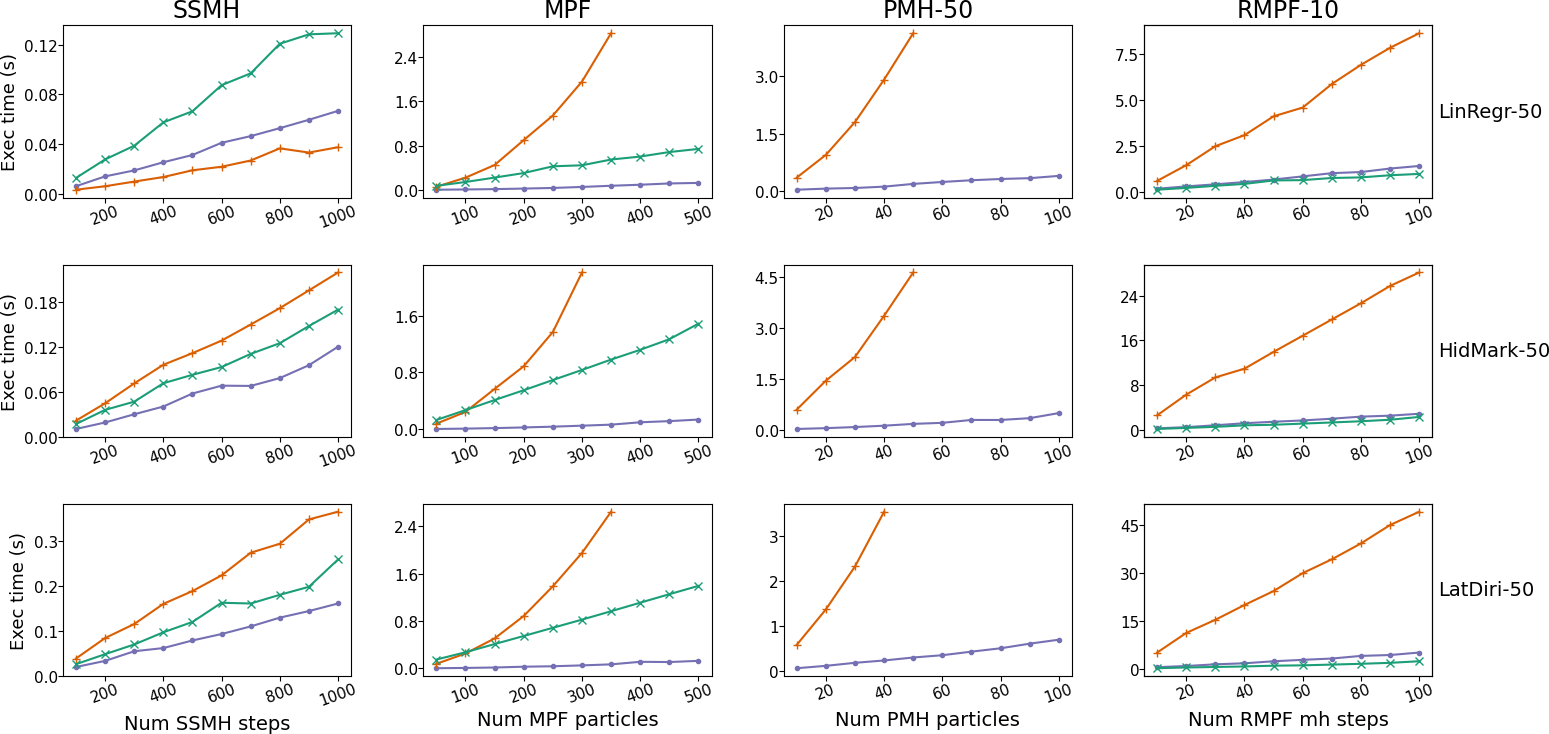}

    \caption{Execution times of inference algorithms (top) with varying number of algorithm iterations or particles. The right-hand axis fixes the number of observations. PMH-50 indicates 50 MH updates that vary in the number of particles, and RMPF-10 indicates 10 particles that vary in the number of MH updates.}
    \label{fig:benchmarks-inference}
    \vspace*{0.3cm}
  \end{subfigure}
  \begin{subfigure}{\textwidth}
    \vspace*{0.2cm}
    \centering
    \includegraphics[width=\textwidth]{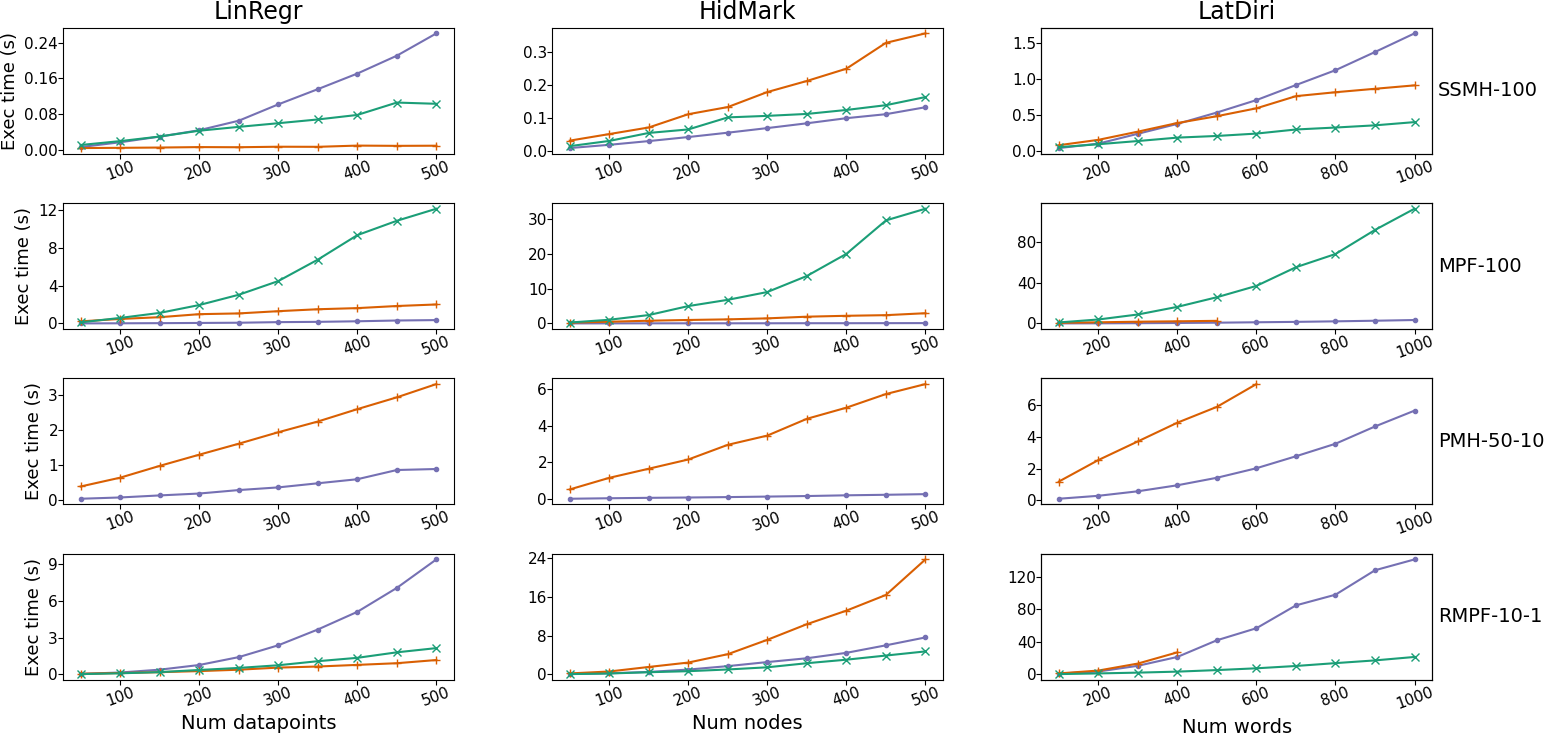}

    \caption{Execution times of inference algorithms (right) with varying number of observations. The right-hand axis fixes the number of algorithm iterations or particles. PMH-50-10 indicates 50 MH updates that use 10 particles; RMPF-10-1 indicates 10 particles that use 1 MH update.}
    \label{fig:benchmarks-model}
  \end{subfigure}
    \vspace*{0.1cm}
  \caption{Performance comparison of our system, ProbFX, with MonadBayes and Gen in terms of mean execution times. The number of executions per mean is left to the control of the benchmarking suites, \href{https://hackage.haskell.org/package/criterion}{Criterion} (Haskell) and \href{https://github.com/JuliaCI/BenchmarkTools.jl}{BenchmarkTools.jl} (Julia). Truncated line plots indicate an algorithm being killed early by the host machine for certain benchmark parameters. Missing line plots indicate an algorithm not being readily implemented in the system.}
\end{figure*}

Before considering how well our approach achieves the goals set out in \secref{background},
we consider how practical it is for actually running inference.
This section shows that our implementation is capable of competing with real-world probabilistic
programming systems, suggesting that the
choice of algebraic effects as a foundation does not imply a compromise on performance.
We compare with two state-of-the-art systems designed
with programmable inference as an explicit goal: MonadBayes~\footnote{\href{https://github.com/tweag/monad-bayes}{github.com/tweag/monad-bayes}} \cite{monadbayes},
a Haskell library that uses a monad transformer effect system, and Gen~\footnote{\href{https://github.com/probcomp/Gen.jl}{github.com/probcomp/Gen.jl}} \cite{genjl}, an embedded language in Julia.

We compared the mean execution times of four algorithms: Single-Site Metropolis-Hastings (SSMH), Multinomial
Particle Filter (MPF), Particle Metropolis-Hastings (PMH), and Resample-Move Particle Filter (RMPF). Each algorithm is applied across three types of
model: linear regression, hidden Markov model, and Latent Dirichlet allocation. These experiments
were carried out on an Intel Core i7-9700 CPU with 16GB RAM.

On average, we outperform either one or both of the other systems across all algorithms, sometimes
asymptotically or by several orders of magnitude. When varying
the number of iterations performed or particles used by each algorithm in \figref{benchmarks-inference},
our performance scales linearly across all models. Our
performance remains linear when varying the number of observations provided to models in \figref{benchmarks-model}, except for RMPF where, like MonadBayes and Gen, we scale quadratically.

Against MonadBayes, for SSMH we are on average \code{15x} slower for linear regression, and \code{1.8x}
faster for other models. The former result is likely because of the specific linear regression model used,
which varies only in the number of \code{observe} operations, and in contrast to our implementation, their
version of SSMH does not store log weights for individual observations, but instead simply sums over them. For
MPF, PMH, and RMPF, we average faster by \code{27x}, \code{16x}, and \code{4.9x} across all models.
When increasing the number of particles in MPF and PMH, the runtime of MonadBayes scales
quadratically, and the process is killed when more than a moderate number of particles are used. We suspect
this is due to their use of the \hsinline{ListT} monad transformer to represent collections of particles,
which in our experience scales poorly as the size of the transformer stack grows.

Comparing with Gen, we are roughly \code{1.1x} and \code{72x} faster for SSMH and MPF, the latter arising
mainly because Gen's MPF implementation scales quadratically with the number of model observations, and for RMPF,
we are on average \code{2.9x} slower. We do not compare PMH since it is not directly provided in Gen, and so
leave this to future work.

\section{Qualitative Comparison}
\label{sec:eval-modularity}

\newcommand*{\CriterionOne}{Reinterpretable models}
\newcommand*{\CriterionTwo}{Modular, reusable algorithms}

\secref{background} identified two key forms of extensibility central to programmable inference, of which we
have seen several examples in the preceding sections:

\vspace{0.05cm}
\begin{enumerate}[leftmargin=*]
   \item \textbf{\emph{\CriterionOne.}} 
   Different algorithms require custom semantics for how models \code{sample} and \code{observe}, as well as
   fine-grained control over model execution in order to implement essential behaviours like suspended
   particles and tracing. ``Programmability'' here means being able to easily customise how models execute in
   order to derive or adapt inference algorithms.\vspace{0.2cm}
   \item \textbf{\emph{\CriterionTwo.}} Different algorithms from the same broad family implement key behaviours like resampling or proposing differently. ``Programmability'' here means being able to
   plug alternative behaviours into an existing algorithm without reimplementing it from scratch, but also
   being able to define new abstract algorithms that are easily pluggable in this way.
\end{enumerate}
\vspace{0.05cm}

\noindent Given that inference programming is often undertaken by domain experts, for whom the activity may
primarily be a means to an end, programmability matters. Here we look at how programmability is achieved in
existing systems, briefly considering dynamically typed settings in
\secref{eval-modularity:dynamically-typed}, and then turning in more detail to MonadBayes
in \secref{eval-modularity:monadbayes}, the main existing system based on typed effects.


\begin{figure*}
\begin{subfigure}{0.47\textwidth}
   \input{code/qualitative/monad-bayes}
   \vspace{-0.2cm}
   \caption{MonadBayes}
   \label{fig:models-monadbayes}
\end{subfigure}
\hspace{0.6cm}
\begin{subfigure}{0.47\textwidth}
   \input{code/qualitative/ours}
   \vspace{-0.2cm}
   \caption{\OurLibrary}
   \label{fig:models-probfx}
\end{subfigure}
\vspace{-0.25cm}
\caption{Support for reinterpretable models}
\label{fig:reinterpretable-models}
\vspace{-0.25cm}
\end{figure*}

\subsection{Dynamically typed approaches}
\label{sec:eval-modularity:dynamically-typed}

Most programmable inference systems to date have been implemented in dynamic languages. We consider Venture,
Gen, Pyro, and Edward; other mainstream systems like Anglican \cite{anglican} and Turing \cite{turingjl} were
not designed with inference programming in mind.

\vspace{0.2cm}
\textbf{\emph{\CriterionOne . \;}}
\label{sec:eval-modularity:dynamically-typed:criterion-one}
In Venture \cite{venture}, modelling and inference instructions are interleaved, with the inference code
affecting the semantics of preceding modelling code; this is flexible but lacks a clear delination between
model and inference. Gen (in Julia) provides a black-box interface for interacting with models, exposing
capabilities such as simulating and tracing, but the operations are non-programmable (have fixed meanings).
Pyro \cite{pyro} and Edward \cite{moore2018effect} (in Python) are more flexible, relying on a stack of
programmable coroutines that are sequentially invoked by \code{sample} and \code{observe} calls; this has some
flavour of algebraic effects, allowing bespoke semantics for \code{sample} and~\code{observe}, albeit without
a type discipline for tracking effects and associating them to handlers, and requiring global state to
maintain the coroutine stack.

Control over model execution is realised in different ways. For particle stepping, Gen requires the programmer
to manage this themselves, parameterising their model on the number of steps to be executed. In Pyro, the programmer must
implement a method \code{step} for any model they want to execute in this way. Other dynamic languages rely on
continuation-passing-style transformations \cite{anglican, webppl}. Algebraic effects seem to offer a clear
advantage here, providing handlers with access to the continuation and making idioms like stepwise execution
easy to implement in inference code, rather than requiring any changes to models.

\vspace{0.2cm}
\textbf{\emph{\CriterionTwo . \;}} Venture offers a range of high-level inference procedures as reusable
primitives, but new inference primitives must be written in Venture's DSL, which cannot reuse external
inference code. In Gen, Pyro, and Edward, the inference libraries are implemented using regular host-language
functions. While technically reusable, the lack of an effect discipline means these functions tend to mix
arbitrary computation with model interactions, rather than being organised explicitly around the key operations of~the algorithm, making them challenging to reuse in new~contexts.

\vspace{-0.1cm}

\subsection{MonadBayes}
\label{sec:eval-modularity:monadbayes}

MonadBayes is a Haskell library for typed programmable inference based on the Monad Transformer Library (MTL).
MTL is an imperative programming framework that allows the programmer to stack monads, producing a combined
effect consisting of ``layers'' of elementary monadic effects called \emph{monad transformers}
\cite{kiselyov2013extensible}. A given set of monads may be layered in different ways; moreover layers can be
abstract, with their operations defined by a type class. To invoke an operation of a specific abstract monad
\code{m} from the stack, the user (or library) must define how each monad transformer above \code{m} \emph{relays} that
operation call further down the stack. A program written in MTL, whose type is an abstract stack of monad
transformers, determines its semantics by instantiating to a particular concrete stack.

\vspace{0.2cm}
\textbf{\emph{\CriterionOne . \;}}
In MonadBayes, reinterpretable models are provided by MTL's support for abstract monad stacks. The constrained
type \hsinline{(MSamp m, MCond m) => m a} represents a model, where the type constructor \code{m} is an
abstract stack of monad transformers, each providing semantics for sampling (\code{rand}) and observing
(\code{score}) by implementing the type classes \hsinline{MSamp} and \hsinline{MCond} in
\figref{models-monadbayes}. Following the usual MTL pattern, each concrete monad must either give a concrete
behaviour for \code{rand} and/or \code{score}, or relay that operation to a monad further down the stack.
For example, the \hsinline{Weighted m} monad is for weighting a model \code{m}; it updates a stored weight
when observing with \code{score}, but simply delegates any calls to \code{rand} to its contained monad
\code{m}, using \code{lift}. The analogue of \hsinline{MSamp} and \hsinline{MCond} in our library are the
concrete datatypes \hsinline{Sample} and \hsinline{Observe} in \figref{models-probfx}, whose operations are
also abstract (now as data constructors), but with semantics given by effect handlers rather than class
instances; the counterpart to the \hsinline{Weighted m} monad is the \code{likelihood} handler which
interprets \hsinline{Observe} to accumulate a weight. The analogue of relaying comes ``for free'' in the
algebraic effects implementation, via \code{handleWith} (\secref{background}).

While monad transformers are both compositional and type-safe, the network of relaying that arises in
MonadBayes is non-trivial. More than one concrete monad in the stack may provide \code{sample} and
\code{observe} behaviours (such as \hsinline{Traced} in \figref{models-monadbayes}, which recursively applies
\code{rand} and \code{score} to its components); others may opt \emph{not} to relay. As relaying is carried
out implicitly, via type class resolution, the eventual runtime behaviour of a model may not be obvious. With
algebraic effects, the correspondence between operations and their semantics is usually more evident, in the
form of handlers, such as the \code{reuseTrace} and \code{likelihood} handlers in \figref{models-probfx} which
provide semantics for \hsinline{Sample} and \hsinline{Observe}.

For control over model execution e.g. for particle stepping, MonadBayes requires the programmer to use
specific control effects, namely the free monad transformer \hsinline{FreeT} and the \hsinline{Coroutine}
monad. Although model authors are oblivious to this particular detail, inference code can still require a
significant amount of plumbing which can obscure the key operations of the algorithm. Algebraic effects instead
provide access to the continuation in each handler, allowing the advertised effect signature to remain
domain-specific.

\vspace{0.25cm}
\textbf{\emph{\CriterionTwo .}}
The reusable building blocks in MonadBayes are datatypes that implement the type classes \hsinline{MSamp} and
\hsinline{MCond} from \figref{models-monadbayes}, such as \hsinline{Weighted} and \hsinline{Traced}. Inference
algorithms are functions that instantiate a model's type from an abstract stack to a specific sequence of
these datatypes. To illustrate, the (simplified) type of \code{rmpf} in \figref{modular-inference-monadbayes},
read inside-out, instantiates the supplied model to ``list of weighted, traced executions''. This expresses
Resample-Move Particle Filter as a computation that nests Metropolis-Hastings (using \hsinline{Traced}) inside
a particle filter (using \hsinline{ListT} for particles). Conversely, the type of \code{pmh} suggests
that Particle Metropolis-Hastings uses a particle filter inside Metropolis-Hastings. Thus the construction of
inference algorithms out of reusable parts is expressed primarily at the type level: by selecting combinations
of datatypes, one determines the specific sampling and conditioning effects that occur at run-time, and the
order in which they interact.

Algebraic effects are similar in a way: the programmer also selects an ordering of abstract operations when instantiating the effect signature
\code{es} in \hsinline{Comp es a}. However, the operations' semantics are not determined by the effect types themselves, but are given separately by effect handlers. For instance, the algorithm \code{rmpf} in \figref{modular-inference-probfx} is
implemented by~choosing a composition of handlers \code{stepModel$_\text{rmpf}$} for executing the model, plus a handler \code{handleResample$_\text{rmpf}$} for the inference effect. Here, constructing inference algorithms out of reusable parts is expressed mainly at the value level, via effect handler~composition.

While each of the concrete monads in MonadBayes is by itself intuitive, for sophisticated algorithms like
\code{rmpf} the transformer stacks can become unwieldy. To extend an algorithm with a new monad, perhaps with
its own type class operations, requires each existing monad in the stack to provide a corresponding instance, and the
new monad in turn to implement each supported operation in the stack. Thus
programmability comes with a certain cost in terms of the amount of boilerplate required. With algebraic
effects, support for new semantics is often more lightweight, requiring only a new handler to define the
relevant operations. For example by swapping out the \hsinline{Resample} handler in multinomial particle
filter (\secref{multinomial}), we were able to derive several other variants
not discussed in the paper such as \emph{residual} and \emph{systematic} particle filter \cite{doucet2009tutorial}, and
also compose these parts to form other algorithms like Resample-Move Particle Metropolis-Hastings \cite{chopin2013smc2}.

\begin{figure}
   \begin{subfigure}{0.5\textwidth}
     \begin{hslistingsmall}
rmpf :: Traced (Weighted (ListT IO)) a -> ...  |\vspace{0.1cm}|
pmh  :: Weighted (ListT (Traced IO)) a -> ...
     \end{hslistingsmall}%
     \vspace{-0.2cm}
     \caption{MonadBayes}
     \label{fig:modular-inference-monadbayes}
     \vspace{0.3cm}
   \end{subfigure}
   \begin{subfigure}{0.5\textwidth}
     \begin{hslistingsmall}
|rmpf| | =   ||{handleResample$_\text{rmpf}$}||  $\circ$ pfilter ||{stepModel$_\text{rmpf}$}| where
   |stepModel$_\text{rmpf}$  = reuseTrace  $\circ$ advance | |\vspace{0.1cm}|
|pmh| = handlePropose$_\text{im}$  $\circ$ mh execModel$_\text{pmh}$ where
   |execModel$_\text{pmh}$ = handleResample$_\text{mul}$ $\circ$ pfilter stepModel$_\text{pmh}$|
     \end{hslistingsmall}%
     \vspace{-0.2cm}
     \caption{\OurLibrary}
     \label{fig:modular-inference-probfx}
   \end{subfigure}
   \vspace{-0.2cm}
   \caption{Support for inference as modular building blocks}
   \label{fig:modular-inference}
   \vspace{-0.3cm}
\end{figure}
\section{Conclusion and Future Work}
\label{sec:conclusion}

Typed functional languages like Haskell offer a type-safe and compositional foundation for inference
programming. However, the intersection of these paradigms can involve a steep learning curve for individuals
not already well versed~in both. This paper presented a technique based on algebraic~effects and
operations for explicating the core structure of inference algorithms, and effect handlers as an intuitive~and
modular interface for programming them. We used this technique to implement some off-the-shelf algorithms in a modular way.

One area of future work is to explore the existing Haskell support for
automatic differentiation \cite{kmett2021ad, van2022forward} and its interplay with effect handlers, which would enable
modern inference techniques like HMC
\cite{chen2014stochastic} and variational autoencoders \cite{kingma2013auto} that require differentiable models. Another is to formalise some
properties of our library. For example, MonadBayes has modular proofs that ensure each of its monad
transformers correctly produces an ``unbiased sampler'' for inference~\cite{scibior2017denotational}.
It may be possible to transfer the semantics of monad transformers to an algebraic effect setting; for example, \citet{goldstein:msc} explores the modularity of Bayesian inference in Koka Bayes: a prototype probabilistic programming library implemented in Koka \cite{Leijen_2014} which takes advantage of its effect type system.
Finally, we are interested in how effect handlers compare to ``unembedding''~\cite{kazutaka2023embedding} as a technique for embedding abstract probabilistic programs; this may allow regular Haskell-bound variables to be assigned the various non-standard semantics that come with probabilistic languages, e.g. of random variables, or optimisable variables that make use of differentiation.

\bibliographystyle{ACM-Reference-Format}
\bibliography{bibliography}

\end{document}